\documentclass[11pt]{article}
\usepackage[utf8]{inputenc}
\usepackage[english]{babel}
\usepackage[final]{graphicx} 
\usepackage{caption}
\usepackage{geometry}
\usepackage{amsmath, amsfonts, amssymb}

\usepackage[babel]{csquotes}

\usepackage{booktabs}
\usepackage{multirow}
\usepackage{float}

\usepackage{comment}

\usepackage[doi=false,isbn=false, backend=biber, maxbibnames=99, maxcitenames=2, style=authoryear, uniquelist=false, uniquename=false]{biblatex}
\DeclareNameAlias{sortname}{family-given}

\usepackage{hyperref}
\addto\extrasenglish{

}
\begin{document}

\begin{tabular}{p{\textwidth}}
 
\medskip

\medskip

\medskip

\medskip

\begin{center}
\LARGE{\textsc{Systematic comparison of Bayesian basket trial designs with unequal sample sizes and proposal of a new method based on power priors}}
\end{center}

\medskip

\medskip

\begin{center}
\begin{tabular}[h]{cc}
\textbf{Sabrina Schmitt} & \textbf{Lukas Baumann } \\
Institute of Medical Biometry & Institute of Medical Biometry  \\
University of Heidelberg & University of Heidelberg \\
69120 Heidelberg, Germany & 69120 Heidelberg, Germany \\
 & \texttt{baumann@imbi.uni-heidelberg.de} \\
\end{tabular}

\medskip

\medskip

September 16, 2024
\end{center}

\end{tabular}

\begin{abstract}
\label{abstract}
\noindent Basket trials examine the efficacy of an intervention in multiple patient subgroups simultaneously. The division into subgroups, called baskets, is based on matching medical characteristics, which may result in small sample sizes within baskets that are also likely to differ. Sparse data complicate statistical inference. Several Bayesian methods have been proposed in the literature that allow information sharing between baskets to increase statistical power. In this work, we provide a systematic comparison of five different Bayesian basket trial designs when sample sizes differ between baskets. We consider the power prior approach with both known and new weighting methods, a design by Fujikawa et al., as well as models based on Bayesian hierarchical modeling and Bayesian model averaging. The results of our simulation study show a high sensitivity to changing sample sizes for Fujikawa's design and the power prior approach. Limiting the amount of shared information was found to be decisive for the robustness to varying basket sizes. In combination with the power prior approach, this resulted in the best performance and the most reliable detection of an effect of the treatment under investigation and its absence.

\end{abstract}

\section{Introduction}
\label{sec:introduction}

\noindent Basket trials allow the efficacy of a single targeted intervention to be studied in multiple diseases or subtypes of the same disease (\cite{woodcock2017}). They are mostly used in oncology, where all patients enrolled in the trial share the same genetic alteration and the treatment being studied targets that genetic predisposition (\cite{pohl2020}). Based on matching characteristics, patients are divided into different subpopulations, called baskets, which can be thought of as parallel sub-trials investigating the same study drug. Basket trials are mainly used in early drug development and usually conducted as uncontrolled phase II trials with a binary endpoint such as drug response, where the goal is to find those subpopulations (baskets) that respond most effectively to treatment.

\noindent Assigning patients to baskets based on medical factors can result in small sample sizes within subgroups that are also likely to differ. Real-world examples such as the cancer study by \textcite{subbiah2023}, which examines eight baskets with sizes between $1$ and $55$, or the study by \textcite{moroSibilot2019}, which investigates a specific form of lung cancer, illustrate this. The latter contains three baskets with sample sizes between $25$ and $37$.

\noindent  Sparse data complicate statistical inference and subsequent decision making. Stratification, in which all subgroups are analyzed separately, is the first of two main options for analysis. However, this is associated with considerable uncertainty due to small sample sizes, and there is a risk of over-interpreting the results of extreme groups. Furthermore, the power to detect the effect of the intervention in each group is often low. Alternatively, assuming homogeneity, one could pool all baskets and test for an effect in the combined population. However, important differences between groups may be lost, which can lead to incorrect conclusions. The appropriate evaluation method depends on the similarity of the baskets, with the methods just mentioned being suitable only for the two extreme cases. If all groups are heterogeneous, separate analysis is appropriate; if they are homogeneous, pooling is the optimal strategy. In practice, however, it is rarely one or the other; usually some of the baskets are similar, while others are not.

\noindent Bayesian methods allow for information sharing (borrowing) across baskets while accounting for potential heterogeneity. Under the assumption that matching genetic predisposition leads to a similar treatment response, information from all baskets is used to estimate the treatment effect in one basket. Several methods have been proposed in the literature that use information sharing to increase statistical power (\cite{pohl2020}). Most of these designs are fully Bayesian, using e.g. Bayesian model averaging (\cite{psioda2021}) or Bayesian hierarchical modeling (\cite{berry2013, neuenschwander2015}), which can be computationally very demanding. Recently, however, new approaches have been developed that use empirical Bayes methods and achieve comparable performance to fully Bayesian methods, but are computationally much cheaper. Examples are the design by \textcite{fujikawa2020} or the power prior approach originally introduced by \textcite{ibrahim2000} and applied to the basket trial setting by \textcite{baumann2024}.

\noindent Several comparisons of basket trial designs have been published in the literature (e.g. \cite{broglio2022, baumann2024}), but most of them neglect the fact that in clinical practice the sample sizes in the individual baskets usually differ. Therefore, this work addresses the question of which study design is best suited for the evaluation of basket trials when the sample sizes of the individual baskets are unequal. 

\noindent Furthermore, we propose a modification of the power prior design for unequal sample sizes. The power prior design for basket trials incorporates data from all baskets using a weighted likelihood that shares information according to the similarity of the individual baskets. However, if the sample sizes of the individual baskets differ, there is a risk that the information from the small baskets will be overlaid by that of the large baskets during the borrowing process and thus lost. \textcite{ollier2020} presented a power prior approach for integrating historical data in a study that accounts for unequal sample sizes. In addition to the weighting parameter, a second parameter is introduced that limits the amount of information shared based on the sample sizes of the baskets, thus avoiding information overlap. We extend the approach of \textcite{ollier2020} to the setting of basket trials and propose two different ways to share data between baskets. Subsequently, the power prior versions are compared with basket trial designs from the literature in an extensive simulation study, considering different sample size scenarios to cover realistic situations that may occur in clinical practice.

\noindent The article is organized as follows. In \autoref{sec:methods} we describe the setting and the notation used. Then, we present the basket trial designs considered. This will be followed by a description of the simulation study in \autoref{sec:simulationstudy}, and the results will be presented in \autoref{sec:results}. The article will conclude with a discussion (\autoref{sec:discussion}).

\section{Methods}
\label{sec:methods}

\subsection{Setup and Notation}
\label{sec:setup}
We consider uncontrolled, single-stage basket trials with a binary endpoint and $K \geq 2$ baskets. Let $\textbf{R} = (R_1,...,R_K)$ be a random vector describing the number of responses in each basket, with $\textbf{r} = (r_1,...,r_K)$ being the corresponding realizations of $\textbf{R}$. Furthermore, we assume the responses in basket $k$ ($k \in \{1,...,K\}$) to be binomially distributed with $R_k \sim \text{Bin}(n_k,p_k)$, where $n_k$ denotes the sample size and $p_k$ probability of response in basket $k$. The latter is unknown and therefore estimated in practice. The data $D_k = \{r_k, n_k\}$ summarize the observed number of responses and the sample size of basket $k$, and $\textbf{D} = (D_1,...,D_K)$ contains the data of all $K$ baskets.

\subsection{Power Prior Design}
\label{sec:PowerPriorApproach}
\noindent \textcite{ibrahim2000} originally introduced power priors to use data from similar historical studies to define the prior distribution of a current clinical trial. The so-called power prior distribution is specified as the likelihood function of the historical data raised to a power $\omega$ that controls its influence. The approach is based on the assumption that the parameter of interest $\theta$ is the same in both data sets (\cite{banbeta2019}).

\noindent For the concrete construction, let $L(\theta|D_0)$ be the likelihood function of $\theta$ conditional on the historical data $D_0 = \{r_0, n_0\}$, where $r_0$ and $n_0$ describe the corresponding number of responses and sample size. Using an initial prior distribution $\pi_0(\theta)$ for $\theta$ and a scalar power prior parameter $\omega \in [0,1]$, the power prior distribution of $\theta$ of the current study is given by
\begin{align}
\pi(\theta|D_0, \omega) \propto L(\theta|D_0)^\omega \pi_0(\theta) \nonumber.
\end{align}

\noindent The power parameter can be considered either fixed or as a random variable which is assigned its own prior distribution. In this work, fixed parameter values are assumed.

\noindent In the setting of basket trials, for binary data and an initial $\text{Beta}(s_{1,k}, s_{2,k})$ prior for the response probability $p_k$ in basket $k$, this leads to the following posterior distribution:
\begin{align}
    \pi(p_k|\textbf{D}, \boldsymbol\omega) = \text{Beta} \bigg(s_{1,k} + \sum_{i=1}^K \omega_{k,i} \cdot r_i, s_{2,k} + \sum_{i=1}^K \omega_{k,i}(n_i - r_i) \bigg)
    \label{eq:posteriorPowerPriorBasketSetting}
\end{align}

\noindent (\cite{baumann2024}). Therefore, the estimation of the treatment effect in one basket includes the information of all $K$ baskets, weighted by the parameters $\omega_{k,i}$.

\noindent In the following, we will present three different ways to calculate the weights in the setting of basket trials.

\subsubsection{Calibrated Power Prior Weights}
\label{sec:CPP_Design}
\textcite{pan2017} introduced the calibrated power prior (CPP) approach to borrow information from historical studies. However, \textcite{baumann2024} showed how it can be applied in the setting of basket trials. 

\begin{figure}[ht!]
    \centering
    \includegraphics[width = 14cm]{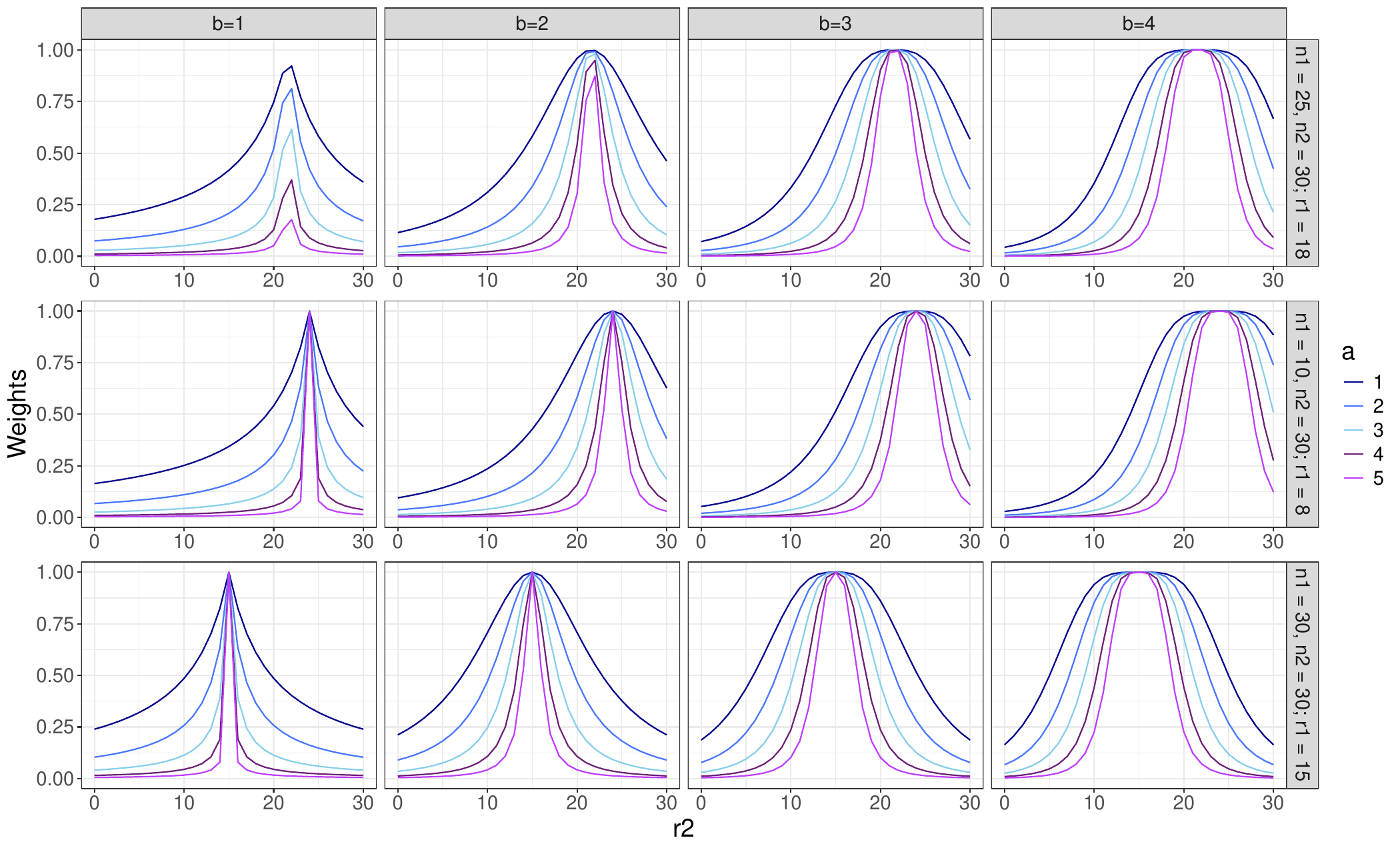}
    \caption[Visualization of the CPP Weights for Different Values of $a$ and $b$.]{CPP weights for different values of $a$ and $b$, considering equal and unequal sample sizes, with both small and large differences.}
    \label{fig:cpp_weights}
\end{figure}

\noindent The Kolmogorov-Smirnov statistic $S_{KS;k,i}$, which corresponds to the absolute rate difference in the case of binary data, is used with the goal of sharing more information the more similar the data sets of the baskets are. \textcite{pan2017} suggested scaling the statistic with $S_{k,i} = \text{max}(n_k, n_i)^{\frac{1}{4}} S_{KS;k,i}$. The weights, which can be interpreted as a pairwise comparison of the response rates of two baskets $k$ and $i$, are computed by
\begin{align}
   \omega_{k,i} = \omega_{k,i}^{CPP} = \frac{1}{1  + \text{exp}(a + b \cdot \text{ln}(S_{k,i}))} \nonumber
\end{align}

\noindent with tuning parameters $a \in \mathbb{R}$ and $b > 0$. It is necessary that $b$ is positive to ensure that larger differences between the data sets result in smaller weights and thus in less information being shared. The influence of the tuning parameters on the weights is illustrated in \autoref{fig:cpp_weights}, considering three different sample size scenarios.

\subsubsection{Adaptive Power Prior Weights}
\label{sec:APP_Design}
Also to incorporate historical data, \textcite{ollier2020} proposed the adaptive power prior (APP) approach, which not only weights the historical information, but also limits the maximum amount of information to be borrowed.

\noindent The adaptive power prior for $\theta$ of the current study is defined by 
\begin{align}
    \pi^{APP}(\theta | D_0, \alpha_0, \gamma)  = \frac{L(\theta|D_0)^{\alpha_0 \, (1-\gamma)} \, \pi_0(\theta)}{\int L(\theta|D_0)^{\alpha_0 \, (1-\gamma)} \, \pi_0(\theta) \, d\theta}, \nonumber
\end{align}

\noindent which is an extension of the power prior approach of \textcite{ibrahim2000}. The original scalar power prior parameter is now split into two parameters, $\alpha_0$ and $\gamma$, which both take values in the interval $[0,1]$. Here, $\alpha_0$ represents the upper limit for the amount of shared information and $\gamma$ denotes the commensurability parameter. 

\noindent Although \textcite{ollier2020} defined their approach to include only a single historical data set in a current study, it can be extended and applied to the basket trial setting using Equation (\ref{eq:posteriorPowerPriorBasketSetting}), defining the weights by  
\begin{align}
   \omega_{k,i} = \alpha_{0;k,i} \, (1-\gamma_{k,i}) \nonumber
\end{align}

\noindent with $\omega_{k,i} = 1$ for $k=i$.

\noindent The two parameters $\alpha_0$ and $\gamma$ are specified for the pairwise comparison of two baskets $k$ and $i$. Considering basket $k$, $\alpha_{0;k,i}$ describes the upper limit for the maximum amount of information borrowed from basket $i$. To ensure that the information from the compared basket $i$ does not superimpose the information in the considered basket $k$, the limit is defined based on the respective sample sizes:
\begin{align}
    \alpha_{0;k,i} = 
    \begin{cases}
        1 & \text{if} \, \, n_k \geq n_i \\
        n_k/n_i & \text{otherwise.} \nonumber
\end{cases}
\label{eq:alpha0}
\end{align}  

\noindent The commensurability parameter ensures that the data sets are not in conflict. To determine the amount of information to be borrowed from basket $i$, the two data sets $D_k$ and $D_i$ are compared using the square of the Hellinger distance:
\begin{equation}
\begin{split}
          d^2(D_i,D_k) = \frac{1}{2} \int \left(\sqrt{\frac{L(\theta|D_k)^{\min(1,\frac{n_i}{n_k})}}{\int L(\theta|D_k)^{\min(1,\frac{n_i}{n_k})} d\theta}}- \sqrt{\frac{L(\theta|D_i)^{\min(1,\frac{n_k}{n_i})}}{\int L(\theta|D_i)^{\min(1,\frac{n_k}{n_i})} d\theta}} \right)^2 \, d\theta, \nonumber
\end{split}
\end{equation}

\noindent where each likelihood has a normalization constant that ensures that it can be seen as a probability distribution. In addition, the exponents automatically downgrade the more accurate likelihood to provide comparability with the less precise likelihood. Note that the general parameter $\theta$ used here corresponds to the response rate $p$ in the basket trial setting. Finally, the commensurability parameter is defined by $\gamma_{k,i} = \sqrt{d^2(D_i, D_k)}$.

\subsubsection{Limited Calibrated Power Prior Weights}
\label{sec:LCPP_Design}
We now propose the limited calibrated power prior (LCPP) approach, which combines the CPP weights, and thus the advantage of tuning with two parameters, with the limitation of the maximum amount of information borrowed between baskets using the $\alpha_0$ parameter from the APP approach.

\noindent Again, the basket-specific response rates are calculated using \autoref{eq:posteriorPowerPriorBasketSetting}, where the weights are defined as follows:
\begin{align}
    \omega_{k,i} = \alpha_{0;k,i} \cdot  \omega_{k,i}^{CPP} \nonumber
\end{align}
 
\noindent with $\omega_{k,i} = 1$ for $k=i$. The parameters $\alpha_{0;k,i}$ and $\omega_{k,i}^{CPP}$ are defined as before.

\subsection{Fujikawa's Design}
\label{sec:fujikawas_design}
The basket trial design proposed by \textcite{fujikawa2020} is very similar to the power prior approach. It also uses basket-specific $\text{Beta}(s_{1,k}, s_{2,k})$ prior distributions for $p_k$ ($k \in \{1,...,K\}$). First, all baskets are modeled individually using a beta-binomial model:
\begin{align}
    \pi (p_k| D_k) = \text{Beta}(s_{1,k} + r_k, \, s_{2,k} + n_k - r_k), \quad \quad k = 1,...,K. \nonumber
\end{align}

\noindent These basket-specific posterior distributions, which do not contain any information from the other baskets, are used in a second step to estimate the final posterior distribution of one basket $k$:
\begin{align}
    \pi(p_k |\textbf{D}) = \text{Beta} \left(\sum_{i = 1}^K \omega_{k,i} \, (s_{1,i} + r_i), \sum_{i = 1}^K \omega_{k,i} \, (s_{2,i} + n_i - r_i) \right). \nonumber
\end{align}

\noindent The information from all baskets is now shared using a weighted sum of the parameters of each basket-specific posterior distribution. There is considerable similarity to the power prior approach (Equation (\ref{eq:posteriorPowerPriorBasketSetting})), with the difference that the prior parameters of all baskets are now within the weighted sum, and thus the basket-specific prior information is also shared. This can be an advantage if the prior distributions are derived from observed previous data. However, if the priors are non-informative, the power prior approach may be more appropriate, since it means that no uncertain information is shared that biases the estimate. 

\noindent The weight $\omega_{k,i}$ represents a measure of the similarity of the posterior distributions of two baskets $k$ and $i$, which determines the amount of information to be borrowed. It is calculated by
\begin{align}
     \omega_{k,i} = 
  \begin{cases}
    \Bigg(1-\text{JSD}\bigg(\pi(p_k|D_k), \pi(p_i|D_i)\bigg)\Bigg)^\varepsilon & \text{if} \, \, \Bigg(1-\text{JSD}\bigg(\pi(p_k|D_k), \pi(p_i|D_i)\bigg)\Bigg)^\varepsilon > \tau \\
    0 & \text{otherwise} \nonumber
  \end{cases}
\end{align}

\noindent with $\omega_{k,i}=1$ for $k=i$. Thus, the amount of information shared between two baskets depends only on the pairwise similarity of their respective basket-specific posterior distributions. 

\noindent To define the weights, the Jensen-Shannon divergence (JSD) is used, which is computed for two probability distributions $W$ and $Q$ by
\begin{align}
   \text{JSD}(W,Q) = \frac{1}{2} \bigg(\text{KLD}(W,M) + \text{KLD}(Q,M)\bigg) \nonumber
\end{align}

\noindent with
\begin{align}
    M=\frac{1}{2} (W + Q) \nonumber
\end{align}

\noindent and $\text{JSD}(W,Q) = \text{JSD}(Q,W)$ (\cite{fujikawa2020}). If the two probability distributions are identical $(W=Q)$, the JSD is 0 (\cite{lin1991}). 

\noindent The JSD is based on the Kullback-Leibler divergence (KLD), defined for two continuous probability densities $W(x)$ and $Q(x)$ as follows (\cite{zhang2023}):
\begin{align}
    \text{KLD}(W,Q) = \int W(x) \, log_2 \, \frac{W(x)}{Q(x)} \, dx. \nonumber
\end{align}

\noindent Note that we use the base $2$ logarithm, as suggested by \textcite{baumann2022}, which results in the JSD taking values between $0$ and $1$ (\cite{lin1991}), and thus $0 \leq \omega_{k,i} \leq 1$. In contrast, the natural logarithm $\text{ln}()$ used by \textcite{fujikawa2020} leads to an upper bound of $\text{ln}(2)$ for the JSD.

\begin{figure}[!ht]
    \centering
    \includegraphics[width = 14cm]{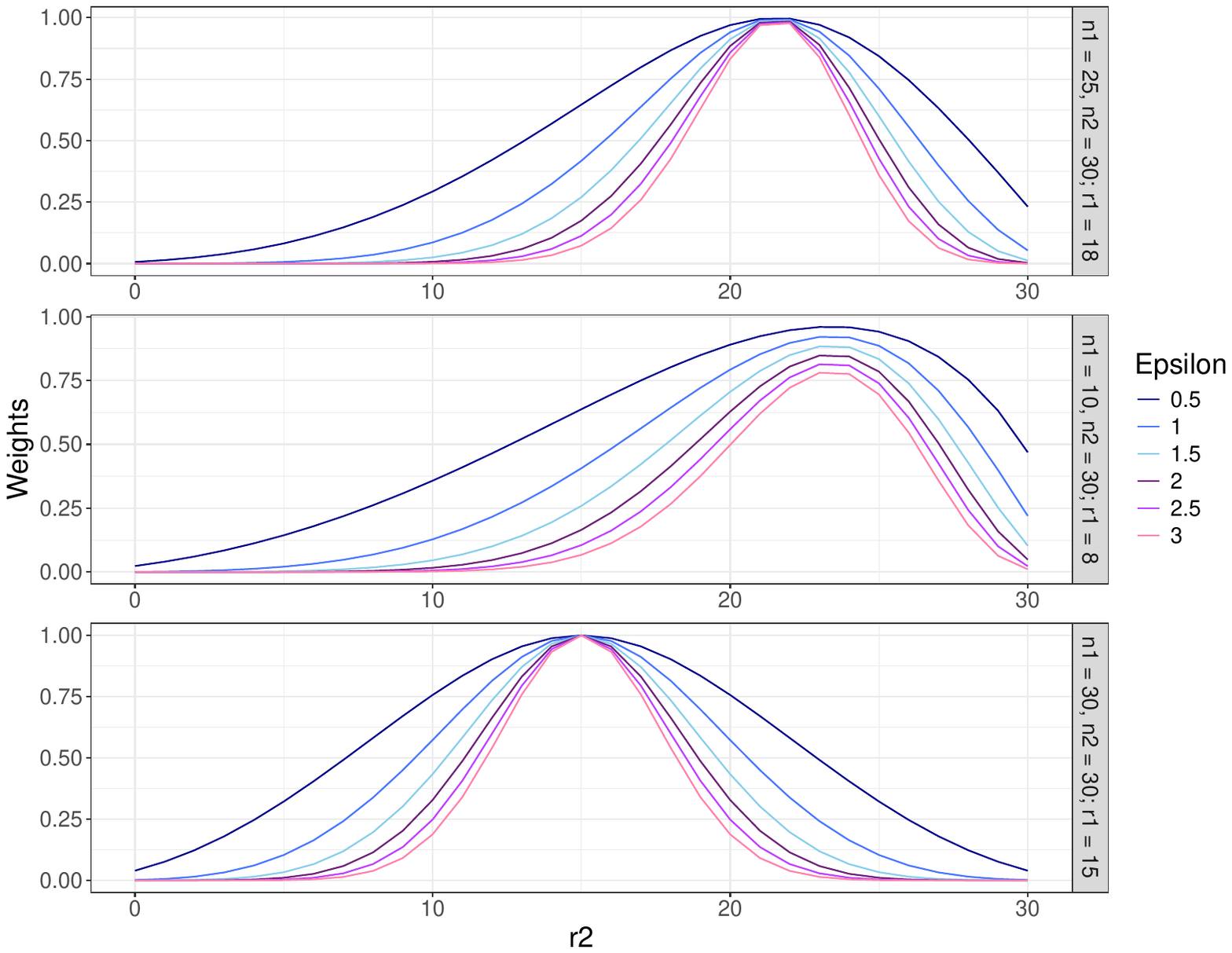}
    \caption[Visualization of the JSD Weights for Different Values of $\varepsilon$.]{JSD weights for different values of $\varepsilon$, considering equal and unequal sample sizes, with both small and large differences.}
    \label{fig:jsd_weights}
\end{figure}

\noindent The JSD weights depend on two tuning parameters $\tau \in [0,1]$ and $\varepsilon > 0$, where the threshold tuning parameter $\tau$ ensures that the baskets have the desired degree of similarity to justify borrowing, and $\varepsilon$ determines how quickly the weights decrease as the similarity of the two posterior distributions declines. \autoref{fig:jsd_weights} visualizes the JSD weights for different values of $\varepsilon$.

\subsection{Bayesian Model Averaging Design}
\label{sec:BMA_Design}
\noindent The Bayesian Model Averaging (BMA) design of \textcite{psioda2021}, first defines all potential models $M_j$ $(j \in \{1,..,J\})$, each describing a distinct classification of the baskets into subsets within which the response rates are identical. For example, considering $K=3$ baskets results in $5$ potential models. In the two extreme cases, all basket-specific response rates either match $(p_1 = p_2 = p_3)$ or differ $(p_1 \neq p_2 \neq p_3)$. Alternatively, only two baskets may have equal response rates: $p_1 = p_2 \neq p_3$, $p_1 \neq p_2 = p_3$, or $p_1 = p_3 \neq p_2$.

\noindent Each potential model $M_j$ $(j \in \{1,..,J\})$ is evaluated separately using the prior $\pi(M_j) \propto \text{exp}(C_j \cdot \psi)$. Where $C_j$ denotes the number of distinct response rates in the model $M_j$ and the tuning parameter $\psi$ weights the potential models. Using $\psi = 0$ assigns the same prior probability to all models, and in the case of $\psi > 0$, those models containing more subsets receive higher prior weights. \textcite{baumann2024} suggested additionally considering $\psi < 0$, which assigns higher prior probabilities to models with fewer subsets. Note that the formula in the model presentation by \textcite{psioda2021} contains a typo, which was corrected in \textcite{baumann2024} after consultation with the first author.

\noindent Finally, inferences about one basket $k$ ($k \in \{1,...,K\}$) can be made using the average of the model-specific posterior probabilities weighted by their posterior model probabilities:
\begin{align}
    Pr(p_k > x | \textbf{D}) = \sum_{j=1}^J Pr(p_k > x|M_j, \textbf{D}) \, \pi(M_j|\textbf{D}). \nonumber
\end{align} 

\noindent The BMA design considers all potential models simultaneously and also incorporates the uncertainties of all models. However, this is computationally quite demanding since all models are always considered, which for a large number of baskets can become a challenge. 

\subsection{Bayesian Hierarchical Model Design}
\label{sec:BHM_Design}
In the Bayesian Hierarchical Model (BHM) design by \textcite{berry2013}, the log-odds of the basket-specific response rates $p_k$ $(k \in \{1,...,K\})$ are modeled with additional adjustments for the respective target rates $p_{targ,k}$:
\begin{align}
    \theta_k = \ln\bigg(\frac{p_k}{1-p_k}\bigg) - \ln\bigg(\frac{p_{targ,k}}{1-p_{targ,k}}\bigg) \nonumber
\end{align}

\noindent with
\begin{align}
    \theta_k \sim N(\mu, \sigma^2). \nonumber
\end{align}

\noindent As a result, similar values of $\theta_k$ across baskets reflect similar treatment effects relative to the respective target values. The fundamental assumption in this approach is that the parameters are exchangeable in their joint distribution.

\noindent The mean $\mu$ and the variance $\sigma^2$ are unknown and each assigned a prior distribution. A normal distribution is again used for the parameter $\mu$. The amount of borrowing is dynamically determined by the data and depends on the similarity of the individual baskets, which is described by the parameter $\sigma^2$. The authors suggest an inverse gamma prior for $\sigma^2$, but it is possible to use other distributions. \textcite{neuenschwander2015} propose to use half-normal, half-Cauchy or half-t as prior distributions for $\sigma$.

\noindent One advantage of hierarchical modeling is shrinkage, where the estimates of all groups are pulled towards the overall mean, thus mitigating random estimation outliers (\cite{berry2013}).

\subsection{EXNEX Design}
\label{sec:EXNEX_Design}
The Exchangeability-nonexchangeability (EXNEX) design by \textcite{neuenschwander2015} is an extension of the exchangeability (EX) approach, which was already used in the BHM design of\break \textcite{berry2013}. Similar to the BHM design, for each basket $k$ the log-odds of the response rate $p_k$ is modeled by
\begin{align}
    \theta_k = \text{ln}\left(\frac{p_k}{1-p_k}\right). \nonumber
\end{align}

\noindent Adding nonexchangeability (NEX) to the full EX model results in two possibilities for each basket-specific parameter $\theta_k$. Each is either
\begin{itemize}
\item EX: with probability $q_k$ exchangeable with parameters of similar baskets. In this case, modeling is done using the normal distribution with the exchangeability parameters $\mu$ and $\sigma$:
\begin{align}
    \theta_k \sim N(\mu, \sigma^2). \nonumber
\end{align}
\item NEX: with probability $1-q_k$ non-exchangeable with any other parameter. For this, basket-specific priors are used 
\begin{align}
    \theta_k \sim N(m_k, v_k), \nonumber
\end{align}
where $m_k$ and $v_k$ are basket-specific parameters that determine the mean and variance of the normal distribution.
\end{itemize}

\noindent The assigned probabilities $q_k$ and $1 - q_k$ are also referred to as weights and should be adapted to the application.

\section{Simulation Study}
\label{sec:simulationstudy}

\noindent A simulation study was conducted using the statistical software R (version 4.3.2), considering the following designs:
\begin{itemize}
\item Power prior design with CPP weights
\item Power prior design with APP weights
\item Power prior design with LCPP weights
\item Fujikawa's design
\item BMA design
\item BHM design
\item EXNEX design
\end{itemize}

\subsection{Simulation Setting}
\label{sec:simulationSetting}
Due to the unlimited possibilities of the number of baskets and their sample sizes, and the aim of a clear comparison, some limitations and specific scenarios have to be chosen. In this simulation study, a total of $N = 100$ patients are assigned to $K = 5$ baskets in three different combinations. This uniform number of observations ensures a meaningful power comparison across scenarios, as larger sample sizes would automatically result in higher power values. Moreover, we decided not to consider baskets with less than $10$ patients to ensure that each basket contains enough information and is not overwritten by the other baskets during borrowing.

\noindent According to \textcite{lengline2021}, the majority of basket trials contain baskets with less than $30$ patients. Therefore, to create realistic scenarios, we set the maximum number of observations per basket to $30$ in the first two scenarios. However, since the literature contains studies where individual baskets have much larger sample sizes (see the study by \cite{subbiah2023}), we additionally defined a third scenario where one basket contains $50$ patients and the others are much smaller, resulting in high variance.

\noindent Initially, a null response rate of $p_0 = 0.15$ was defined and different patterns of actual (true) response rates within the baskets were explored. In order to keep the study manageable, six response rate combinations were selected, including both homogeneous and heterogeneous scenarios. These are partly inspired by the simulation studies of \textcite{broglio2022} and \textcite{baumann2024} and are referred to as patterns. All $18$ simulation scenarios are visualized in \autoref{tab:Simscenarios}.

\begin{table}[ht!]
\centering
\caption[Simulation Scenarios]{Simulation scenarios considered in the comparison study, consisting of the sample sizes $n_k$ and the true response rate $p_k$ in the baskets.}
\begin{tabular}{l l c c c c c c} 
\toprule
Scenario &  & \multicolumn{5}{c}{k} & Pattern\\ \cmidrule{3-7}
         &  & 1 & 2 & 3 & 4 & 5 &            \\
\midrule
  & $p_k$ & \textbf{0.15} & \textbf{0.15} & \textbf{0.15} & \textbf{0.15} & \textbf{0.15} & \multirow{4}{*}{Null}\\ 
1 & $n_k$ & 10 & 15 & 20 & 25 & 30 &                              \\
2 & $n_k$ & 10 & 10 & 25 & 25 & 30 &                              \\ 
3 & $n_k$ & 10 & 10 & 10 & 20 & 50 &                              \\                 
\hline
& $p_k$ & \textbf{0.35} & \textbf{0.35} & \textbf{0.35} & \textbf{0.35} & \textbf{0.35} & \multirow{4}{*}{Alternative}\\                
4 & $n_k$ & 10 & 15 & 20 & 25 & 30 &                              \\
5 & $n_k$ & 10 & 10 & 25 & 25 & 30 &                              \\  
6 & $n_k$ & 10 & 10 & 10 & 20 & 50 &                              \\ 
\hline
  & $p_k$ & \textbf{0.15} & \textbf{0.15} & \textbf{0.25} & \textbf{0.35} & \textbf{0.35} & \multirow{4}{*}{Ascending} \\
7 & $n_k$ & 10 & 15 & 20 & 25 & 30 &                              \\ 
8 & $n_k$ & 10 & 10 & 25 & 25 & 30 &                              \\ 
9 & $n_k$ & 10 & 10 & 10 & 20 & 50 &                              \\ 
\hline
   & $p_k$ & \textbf{0.35} & \textbf{0.35} &\textbf{ 0.25} & \textbf{0.15} & \textbf{0.15} & \multirow{4}{*}{Descending} \\ 
10 & $n_k$ & 10 & 15 & 20 & 25 & 30 &                            \\ 
11 & $n_k$ & 10 & 10 & 25 & 25 & 30 &                             \\ 
12 & $n_k$ & 10 & 10 & 10 & 20 & 50 &                             \\ 
\hline
   & $p_k$ & \textbf{0.15} & \textbf{0.15} & \textbf{0.15} & \textbf{0.15} & \textbf{0.40} & \multirow{4}{*}{Big Good Nugget} \\ 
13 & $n_k$ & 10 & 15 & 20 & 25 & 30 &                             \\ 
14 & $n_k$ & 10 & 10 & 25 & 25 & 30 &                             \\ 
15 & $n_k$ & 10 & 10 & 10 & 20 & 50 &                             \\ 
\hline
& $p_k$ & \textbf{0.40} & \textbf{0.15} & \textbf{0.15} & \textbf{0.15} & \textbf{0.15} & \multirow{4}{*}{Small Good Nugget}\\
16 & $n_k$ & 10 & 15 & 20 & 25 & 30 &                             \\ 
17 & $n_k$ & 10 & 10 & 25 & 25 & 30 &                             \\ 
18 & $n_k$ & 10 & 10 & 10 & 20 & 50 &                             \\ 
\bottomrule
\end{tabular}
\label{tab:Simscenarios}
\end{table}

\subsection{Implementation}
\label{sec:implementation}
Data generation, implementation of all models, their parameter tuning and evaluation were mainly conducted using the R package \textit{basksim} (\url{https://github.com/sabrinaschmitt/basksim}), which also calls functions from the packages \textit{bmabasket} (\cite{bmabasket}) and \textit{bhmbasket} (\cite{bhmbasket}). 

\noindent For each of the $18$ simulation scenarios, data sets have been generated with binomially distributed response rates for all $K = 5$ baskets. Each data set consists of $10,000$ simulation runs. To ensure comparability, all designs were evaluated using the same data sets. 

\noindent In addition, the designs BHM and EXNEX require MCMC sampling to approximate the posterior distributions. We used $10,000$ samples with the first third of the iterations discarded as a burn-in period. 

\noindent The computations for parameter tuning and evaluation were parallelized using the R package \textit{doFuture} and the accuracy of the optimization of $\lambda$ was three decimals for all models.

\noindent See \url{https://github.com/sabrinaschmitt/simulation-study-basksim-manuscript.git} for the\break complete code used in the simulation study.

\subsection{Decision Making}
\label{sec:decisionMaking}
\noindent To decide whether a basket is active, i.e. whether the treatment under investigation is classified as effective in the basket, a family of hypotheses is specified:
\begin{align}
    H_0: p_k \leq p_0 \quad \text{vs} \quad H_1: p_k>p_0,  \qquad k = 1,...,K. \nonumber
\end{align}

\noindent Whether basket $k$ is considered active is based on the posterior distribution of $p_k$ and a probability threshold $\lambda \in (0,1)$. \textcite{fujikawa2020} suggested the final decision rule $Pr(p_k > p_0 | \textbf{D}) \geq \lambda$, which we also used for the power prior design. The decision-making in both the BMA design and the BHM design is based on $Pr(p_k > p_0 | \textbf{D}) > \lambda$. This has been adopted for the EXNEX design.

\subsection{Parameter Tuning}
\label{sec:parameterTuning}
Most of the designs considered require prior distributions and tuning parameters that affect the amount of information shared between baskets and thus the model performance. Therefore, in order to evaluate their best possible performance, all methods were first individually optimized using the generated data sets by identifying the combination of parameter values that maximizes the mean expected number of correct decisions (ECD) averaged over all patterns. In this context, a correct decision is defined as the conclusion of an activity if the basket's true response rate $p_k$ is above the null response rate $p_0$, or no activity if $p_k \leq p_0$. Thus, the ECD describes the expected number of baskets correctly classified as active or inactive, ranging from $0$ to $K$.

\noindent The parameter tuning procedure is based on the simulation study of \textcite{broglio2022} and was performed separately for all three sample size scenarios. First, we defined sets of prior and tuning parameter values for all designs. Then, in a first optimization step, $\lambda$, i.e. the probability threshold used to decide whether the basket under consideration is classified as active (see \autoref{sec:decisionMaking}), was optimized for each model and each combination of possible parameter values. The threshold parameter was selected so that the one-sided family wise error rate (FWER) $\alpha = 0.05$ was maintained under the global null hypothesis. Then, for each model, the ECD was calculated for all patterns using the optimal $\lambda$ values, and the parameter value combination that yielded the highest ECD averaged over all patterns was selected and used in the subsequent model comparison.

\noindent The selection of potential parameter values is mostly equivalent to that of \textcite{baumann2024}. As in \textcite{broglio2022}, some parameters were held fixed for all models, as they were assumed to have little influence on borrowing and thus the impact on model performance would be small. For each of the remaining parameters, a number of possible values were defined, which are now presented in detail. 

\noindent In the power prior design, Fujikawa's design, and the BMA design, we used a fixed $\text{Beta}(1,1)$ prior for the response rates in all baskets.

\noindent The two power prior design versions with CPP and LCPP weights include the two tuning parameters $a$ and $b$. Both were optimized in the range from $0.5$ to $5$ in $0.5$ steps, separately for the two weight options. \textcite{baumann2024} only considered the interval $[0,3]$. However, since the upper limit of $3$ was reached several times in this simulation study, the optimization range was extended in order to obtain the best possible model performances.

\noindent The design version using APP weights does not contain any tuning parameters.

\noindent For Fujikawa's design, the tuning parameters $\varepsilon$ and $\tau$ were optimized in the intervals $[0.5,3]$ and $[0,0.5]$, respectively, with $0.5$ steps considered for $\varepsilon$ and $0.1$ steps for $\tau$.

\noindent In addition to the beta prior distribution, the BMA design contains the tuning parameter $\psi$, for which values in $0.5$ steps from $-4$ to $4$ were considered in the optimization process.

\noindent The BHM design has a fixed prior for $\mu$, which was set to $N(-1.1156,100^2)$. This is a weakly informative prior distribution whose mean corresponds to the null hypothesis when $p_0 = 0.15$ and a common target response rate of $0.35$ are considered. The \textit{bhmbasket} package uses a half-normal distribution for the prior of $\sigma$, whose scaling parameter $\phi$ has been optimized using $8$ equidistant values in the interval $[0.125,2]$.

\noindent In the EX part of the EXNEX design, the log-odds of the response rates are modeled with a normal distribution. For their mean, the fixed prior $N(-1.7346,100^2)$ is used. The mean of this weak prior distribution also corresponds to the null hypothesis. However, since the EXNEX model does not adjust for target response rates, the mean of the distribution differs from that of the BHM design. A half-normal distribution was again used for the tuning parameter $\sigma$ and its scaling parameter $\phi$ was optimized with the same values as in the BHM design. The NEX part requires additional basket-specific prior distributions, which were also set to $N(-1.7346,100^2)$ across all baskets. The weighting parameters $q_k$ $(k \in \{1,...,K\})$ were also defined equally for all baskets, which is why the simplified notation $q$ is used in the following. Values from $0.1$ to $0.9$ were investigated in $0.1$ steps.

\subsection{Evaluation}
\label{sec:evaluation}
Separate comparisons were conducted for the three sample size scenarios using the optimal tuning parameter values of each design. We evaluated the performance of each model within each response rate pattern using frequentist operating characteristics, even though all designs considered are Bayesian.

\noindent The ECD was evaluated both within patterns and averaged across patterns to assess the overall performance of the designs. In addition, the basket-specific type I error rate (TOER) and power were considered, which are summarized under the term rejection rate. Finally, the FWER and bias were also taken into account.

\noindent In the tuning process, the probability threshold $\lambda$ was chosen to protect the one-sided FWER at $0.05$ under the global null hypothesis, which leads to increased TOERs in those patterns that contain both active and inactive baskets. Since high power values in the active baskets are often associated with high TOERs in the inactive baskets, power is not the optimal comparison parameter in this case. Therefore, we used the mean ECD as the main comparison parameter instead.

\section{Results}
\label{sec:results}
The initial tuning process resulted in different optimal tuning parameter values for the three sample size scenarios, listed in \autoref{tab:optTunParameter}. 

\begin{table}[ht!]
\centering
\caption[Optimal Tuning Parameter Values.]{Optimal tuning parameter values.}
\begin{tabular}{lcccc}
  \toprule
  Design & Parameter &  \multicolumn{3}{c}{Scenario} \\ \cmidrule{3-5}
  & & Linear & Grouped & High Variance \\ 
  \hline
  \multirow[t]{2}{*}{CPP} & a  & 4 & 4 & 4\\ 
                          & b  & 4.5 & 4.5 & 4\\ 
  \multirow[t]{2}{*}{LCPP} & a  & 3 & 3 & 2.5\\ 
                           & b  & 4 & 4.5 & 5\\ 
  \multirow[t]{2}{*}{Fujikawa} &  $\varepsilon$ & 1.5 & 1.5 & 2.5\\ 
                               &  $\tau$ &  0.2 & 0 & 0.2\\ 
  BMA & $\psi$ & -2 & -2 & -2\\ 
  BHM & $\phi$ & 0.661 & 0.661 & 0.661 \\ 
  \multirow[t]{2}{*}{EXNEX} & $\phi$ & 0.661 & 0.661 & 0.661\\
                            & $q$ & 0.9 & 0.9 & 0.8\\
   \bottomrule
\end{tabular}
\label{tab:optTunParameter}
\end{table}

\noindent \autoref{tab:ecd} shows the ECD achieved by the individual designs in the considered sample size scenarios: Linear, Grouped and High Variance. For each design, the results are given for each pattern and also averaged over all patterns (Mean), using the optimal tuning parameter values for each method. 

\noindent The patterns Big Good Nugget (BGN) and Small Good Nugget (SGN) are abbreviated in the results tables.

\begin{table}[H]
\centering
 \caption[ECD]{ECD in the three sample size scenarios: Linear, Grouped and High Variance. Within each scenario, the best result per column is highlighted in bold.}
{\small
\begin{tabular}{llccccccc}
  \toprule
 Scenario & Design & Null & Alternative & Ascending & Descending & BGN & SGN & Mean\\ 
  \hline
\multirow[c]{7}{*}{Linear}  
     & CPP & 4.914 & \textbf{4.643} & 3.725 & 3.232 & 4.356 & \textbf{4.298} & 4.195 \\ 
     & APP & 4.925 & 4.618 & 3.906 & 3.163 & 4.511 & 4.110 & 4.205 \\ 
     & LCPP & 4.921 & 4.633 & \textbf{3.949} & 3.198 & 4.475 & 4.259 & \textbf{4.239} \\ 
     & Fujikawa & 4.908 & 4.615 & 3.627 & \textbf{3.301} & 4.317 & 4.175 & 4.157 \\
     & BMA & 4.914 & 4.548 & 3.790 & 2.997 & 4.543 & 4.112 & 4.151 \\ 
     & BHM & \textbf{4.929} & 4.548 & 3.783 & 3.037 & \textbf{4.588} & 4.101 & 4.164 \\ 
     & EXNEX & \textbf{4.929} & 4.546 & 3.818 & 3.039 & 4.583 & 4.106 & 4.170 \\ 
\hline
\multirow[c]{7}{*}{Grouped}  
     & CPP & 4.919 & 4.609 & 3.717 & 3.097 & 4.347 & \textbf{4.269} & 4.160 \\ 
     & APP & 4.927 & 4.547 & 4.031 & 3.021 & 4.519 & 4.114 & 4.193 \\ 
     & LCPP & 4.925 & 4.593 & \textbf{4.147} & 2.997 & 4.435 & 4.251 & \textbf{4.225} \\ 
     & Fujikawa & 4.900 & \textbf{4.673} & 3.568 & \textbf{3.159} & 4.242 & 4.105 & 4.108 \\
     & BMA & 4.913 & 4.527 & 3.827 & 2.890 & 4.547 & 4.113 & 4.136\\ 
     & BHM & 4.932 & 4.507 & 3.837 & 2.909 & 4.590 & 4.104 & 4.146 \\ 
     & EXNEX & \textbf{4.933} & 4.485 & 3.867 & 2.882 & \textbf{4.598} & 4.106 & 4.145 \\  
\hline
\multirow[c]{7}{*}{High Variance} 
     & CPP & 4.921 & 4.459 & 3.724 & 2.974 & 4.310 & \textbf{4.246} & 4.106 \\ 
     & APP & 4.928 & 4.385 & 3.739 & 3.031 & 4.662 & 4.129 & 4.146 \\ 
     & LCPP & 4.918 & \textbf{4.557} & 3.697 & \textbf{3.143} & 4.618 & 4.131 & \textbf{4.177} \\ 
     & Fujikawa & 4.910 & 4.367 & 3.646 & 3.138 & 4.290 & 4.178 & 4.088 \\ 
     & BMA & 4.913 & 4.431 & 3.759 & 2.667 & 4.661 & 4.121 & 4.092 \\ 
     & BHM & 4.928 & 4.421 & 3.725 & 2.688 & \textbf{4.695} & 4.121 & 4.096 \\ 
     & EXNEX & \textbf{4.930} & 4.386 & \textbf{3.770} & 2.670 & 4.673 & 4.130 & 4.093 \\ 
   \bottomrule
\end{tabular}
}
   \label{tab:ecd}
\end{table}

\noindent In all three scenarios, the highest ECDs are achieved when all baskets are either active or inactive, since borrowing works best in these patterns. When the activity status of one basket differs from that of the other baskets (SGN and BGN), all designs record fewer correct classifications. It makes a difference whether the individual basket is active or inactive, since in all three scenarios, the BGN pattern, where the largest basket shows activity, has higher values than the SGN pattern, where only the smallest basket is active. The lowest results are seen in the highly heterogeneous patterns Ascending and Descending, where less information can be borrowed between the individual baskets due to the high heterogeneity. While in all other patterns on average more than $4$ baskets are correctly classified, the ECDs are now mostly less than 4, with the exception of the power prior design with APP and LCPP weights in the grouped scenario. In the Descending pattern, where the active baskets are small and the inactive baskets are large, even values below $3$ are recorded, with the number of correctly classified baskets decreasing as the variance of the sample sizes increases. Overall, the Ascending and Descending patterns show the greatest variance between the results of the different designs.

\noindent In comparison, no design always achieves the highest ECD; rather, the performance of the individual models varies between the different response rate patterns and also between the sample size scenarios. However, Fujikawa's design performs worst in all three scenarios in the Null, Ascending, and BGN patterns. Both Ascending and BGN have in common that the active baskets are large and the inactive baskets are small. Furthermore, the design shows a poorer performance as the variance in the sample sizes of the individual baskets increases, while the other methods show no tendency. No version of the power prior design ever shows the least ECD.

\noindent However, when looking at the number of correct decisions made by the designs, averaged across all patterns, a clear trend emerges. In all three sample size scenarios, the three versions of the power prior approach perform best, with the LCPP weighting achieving the most correct classifications, followed by the APP and CPP weights. The two hierarchical models record the next highest values, with the EXNEX design performing slightly better in the linear scenario and the BHM performing better in the grouped and high variance sample size scenarios. However, the differences between the designs are small. In all three scenarios, the poorest results are observed for the BMA and Fujikawa's designs, with the latter performing slightly better in the linear case and the BMA design in the remaining two patterns.

\begin{table}[p]
\centering
   \caption[Rejection Rates and FWERs for the grouped Scenario.]{Basket-wise rejection rates and FWERs for the grouped scenario. For each pattern, results are shown for all designs. Within patterns, the active baskets are highlighted in bold.}
{\small
\begin{tabular}{lccccccc}
  \toprule
\multirow{2}{*}{Pattern} & \multirow{2}{*}{Design} & Basket 1 & Basket 2 & Basket 3 & Basket 4 & Basket 5 & \multirow{2}{*}{FWER} \\ 
& & $n_1 = 10$ & $n_2 = 10$ & $n_3 = 25$ & $n_4 = 25$ & $n_5 = 30$ & \\
  \hline
\multirow{7}{*}{Null}  
                      & CPP & 0.020  & 0.019  & 0.014  & 0.014  & 0.013  & 0.048  \\ 
                      & APP & 0.008  & 0.009  & 0.020  & 0.018  & 0.018  & 0.049  \\ 
                      & LCPP & 0.015  & 0.013  & 0.015  & 0.015  & 0.016  & 0.050  \\ 
                      & Fujikawa & 0.018  & 0.019  & 0.022  & 0.022  & 0.020  & 0.048  \\
                      & BMA & 0.012  & 0.012  & 0.021  & 0.020  & 0.021  & 0.049  \\ 
                      & BHM & 0.007  & 0.007  & 0.019  & 0.017  & 0.018  & 0.052 \\ 
                      & EXNEX & 0.007  & 0.007  & 0.018  & 0.017  & 0.017  & 0.049  \\ 
   \hline
 \multirow{7}{*}{Alternative}  
                              & CPP & \textbf{0.886} & \textbf{0.893} & \textbf{0.942} & \textbf{0.942} & \textbf{0.946} & 0.000 \\ 
                              & APP & \textbf{0.832} & \textbf{0.839} & \textbf{0.955} & \textbf{0.954} & \textbf{0.967} & 0.000 \\ 
                              & LCPP & \textbf{0.838} & \textbf{0.843} & \textbf{0.969} & \textbf{0.968} & \textbf{0.975} & 0.000 \\ 
                              & Fujikawa & \textbf{0.915} & \textbf{0.918} & \textbf{0.945} & \textbf{0.946} & \textbf{0.950} & 0.000 \\
                              & BMA & \textbf{0.855} & \textbf{0.859} & \textbf{0.932} & \textbf{0.933} & \textbf{0.948} & 0.000 \\ 
                              & BHM & \textbf{0.839} & \textbf{0.846} & \textbf{0.935} & \textbf{0.937} & \textbf{0.952} & 0.000 \\ 
                              & EXNEX & \textbf{0.829} & \textbf{0.836} & \textbf{0.932} & \textbf{0.934} & \textbf{0.951} & 0.000 \\ 
    \hline
 \multirow{7}{*}{Ascending} 
                            & CPP & 0.319 & 0.322 & \textbf{0.621} & \textbf{0.863} & \textbf{0.874} & 0.487 \\ 
                            & APP & 0.160 & 0.164 & \textbf{0.609} & \textbf{0.859} & \textbf{0.887} & 0.248 \\ 
                            & LCPP & 0.165 & 0.167 & \textbf{0.699} & \textbf{0.876} & \textbf{0.904} & 0.239 \\ 
                            & Fujikawa & 0.406 & 0.405 & \textbf{0.621} & \textbf{0.876} & \textbf{0.882} & 0.608 \\ 
                            & BMA & 0.218 & 0.217 & \textbf{0.533} & \textbf{0.848} & \textbf{0.881} & 0.302 \\ 
                            & BHM & 0.184 & 0.186 & \textbf{0.510} & \textbf{0.829} & \textbf{0.868} & 0.277 \\ 
                            & EXNEX & 0.182 & 0.183 & \textbf{0.524} & \textbf{0.834} & \textbf{0.875} & 0.279 \\ 
    \hline
 \multirow{7}{*}{Descending} 
                             & CPP & \textbf{0.495} & \textbf{0.494} & \textbf{0.274} & 0.084 & 0.082 & 0.124 \\ 
                             & APP & \textbf{0.443} & \textbf{0.448} & \textbf{0.389} & 0.129 & 0.129 & 0.201 \\ 
                             & LCPP & \textbf{0.472} & \textbf{0.475} & \textbf{0.305} & 0.131 & 0.123 & 0.176 \\ 
                             & Fujikawa & \textbf{0.514} & \textbf{0.514} & \textbf{0.352} & 0.114 & 0.106 & 0.174 \\ 
                             & BMA & \textbf{0.388} & \textbf{0.395} & \textbf{0.334} & 0.111 & 0.116 & 0.167 \\ 
                             & BHM & \textbf{0.374} & \textbf{0.378} & \textbf{0.360} & 0.098 & 0.105 & 0.168 \\ 
                             & EXNEX & \textbf{0.362} & \textbf{0.365} & \textbf{0.348} & 0.092 & 0.100 & 0.161 \\
    \hline
 \multirow{7}{*}{BGN}  
                      & CPP & 0.137 & 0.133 & 0.086 & 0.081 & \textbf{0.784} & 0.296 \\ 
                      & APP & 0.060 & 0.057 & 0.083 & 0.081 & \textbf{0.800} & 0.190 \\ 
                      & LCPP & 0.061 & 0.057 & 0.104 & 0.099 & \textbf{0.756} & 0.207 \\ 
                      & Fujikawa & 0.185 & 0.183 & 0.096 & 0.091 & \textbf{0.797} & 0.346 \\
                      & BMA & 0.059 & 0.055 & 0.068 & 0.064 & \textbf{0.793} & 0.166 \\ 
                      & BHM & 0.050 & 0.048  & 0.066 & 0.062 & \textbf{0.816} & 0.162 \\ 
                      & EXNEX & 0.047  & 0.044  & 0.065 & 0.061 & \textbf{0.815} & 0.157 \\
    \hline
 \multirow{7}{*}{SGN}  
                      & CPP & \textbf{0.386} & 0.046  & 0.024  & 0.024  & 0.024  & 0.074 \\ 
                      & APP & \textbf{0.276} & 0.042  & 0.041  & 0.042  & 0.039  & 0.108 \\ 
                      & LCPP & \textbf{0.389} & 0.047  & 0.029  & 0.030  & 0.031  & 0.086 \\
                      & Fujikawa & \textbf{0.269} & 0.056 & 0.037  & 0.037  & 0.034  & 0.097 \\
                      & BMA & \textbf{0.261} & 0.032  & 0.037  & 0.038  & 0.041  & 0.091 \\ 
                      & BHM & \textbf{0.243} & 0.024  & 0.036  & 0.037  & 0.040  & 0.101 \\ 
                      & EXNEX & \textbf{0.239} & 0.023  & 0.034  & 0.036  & 0.038  & 0.097 \\ 
   \bottomrule
   \end{tabular}
   }
   \label{tab:rr_grouped}
\end{table}

\newpage
\noindent \autoref{tab:rr_grouped} shows the basket-specific rejection rates and the FWERs in the grouped sample size scenario. See the \nameref{Supplementary Material} for the results of the scenarios Linear and High Variance.

\noindent Larger sample sizes and more active baskets increase the power since more information is available. Higher true response rates within the baskets lead to further improvement. Overall, the highest values are observed in the Alternative pattern, where all baskets show activity and thus a lot of borrowing is enabled. Fujikawa's design and the power prior design with CPP weights perform best in the small baskets, while using LCPP and APP weights gives the highest power in the large baskets. In the heterogeneous patterns with multiple active baskets (Ascending and Descending), Fujikawa's design and the power prior design versions predominantly achieve good power values in the design comparison. However, when only one basket shows activity, both hierarchical models achieve good values if the single active basket is large (BGN), and using LCPP and APP weights in the power prior approach results in the highest power for a small sized single active basket (SGN).

\noindent Controlling the TOER was difficult because $\lambda$ was chosen in the tuning process to protect the one-sided FWER at $\alpha = 0.05$ under the global null hypothesis. However, this resulted in increased error rates in the heterogeneous patterns, most of which exceed the $0.05$ significance level. 

\noindent Overall, lower and therefore better TOERs are recorded with more inactive baskets, which further improve with higher sample sizes due to the larger amount of information available for borrowing. Within patterns, two distinct trends emerge. 

\noindent Fujikawa's design and the power prior design with CPP weights record very high TOERs in the small baskets of the Ascending and BGN patterns, well above those of the other designs. In the Ascending pattern, Fujikawa's design has error rates over $40 \%$ and using CPP weights results in rates over $31 \%$, while the values of the other designs range between $16 \%$ and $22 \%$. With over $18 \%$ (Fujikawa) and over $13 \%$ (CPP), their error rates improve slightly in the BGN pattern, but still exceed those of the other methods, which are between $4$ and $6$ percent. The reason for the high TOERs is the lack of restriction on the amount of information borrowed from larger baskets. Therefore, a lot of unsuitable information from the large active baskets is included in the response rate estimation of the small inactive baskets, potentially overlaying the information available there and causing an overestimation. Overestimated response rates incorrectly lead to a conclusion of activity, which increases the\break TOER. 

\noindent Due to the additional sharing of prior information, the TOERs of Fujikawa's design exceed those of the power prior design with CPP weights. We used a $\text{Beta}(1,1)$ prior in all baskets, which corresponds to a $50 \%$ rate. Thus, the estimation of each basket-specific response rate includes the $50 \%$ priors from all five baskets (regardless of their activity status) in addition to the actual observed information, which further increases the estimate. 

\noindent With more inactive baskets and larger sample sizes, both designs show improvement, as the fewer and smaller active baskets have less impact on the estimation. In fact, in the large baskets, the use of CPP weights results in the lowest error rates.

\noindent In contrast, the remaining designs show TOER inflation with larger sample sizes. Using LCPP and APP weights (instead of CPP) in the power prior design reduces the TOER in the small baskets of the Ascending and BGN patterns. Both LCPP and APP weights limit the amount of information borrowed from larger baskets based on the ratio of the corresponding sample sizes. Therefore, less mismatching information from the large active baskets biases the response rate estimation in the small inactive baskets. As the sample size increases, both designs show an increase in TOER as more information is shared from the active baskets. When considering the fifth (inactive) basket in the Descending pattern, the effect of the quantity limiting parameter is absent because all remaining baskets from which information could potentially be borrowed are smaller. Nevertheless, the error rates of both designs exceed those obtained using the CPP weights. Although the LCPP weights include the CPP weights, the tuning process yielded different values for the tuning parameters $a$ and $b$, which affect the borrowing and thus the model performance. The APP weights, on the other hand, are based on a different similarity measure and do not contain any tuning parameters, which explains the differing results.

\noindent Both hierarchical models show very similar results, which can be explained by their similarity to each other. This was further increased by the tuning process, that provided identical prior distributions for the exchangeability parameter $\sigma$ (present in both models) in all sample size scenarios, which is responsible for borrowing. In addition, large weights $q$ were used for the exchangeability part in the EXNEX design, further increasing the similarity. Overall, both designs achieve low TOERs, especially in the small baskets.

\noindent In general, the differences between baskets of the same size are marginal for TOER and power, and thus cannot be distinguished from simulation errors.

\noindent The trends observed for the TOER are also reflected in the FWER. The one-sided FWER of $\alpha = 0.05$ is mostly maintained in the Null pattern, with a small exceedance of $2 \%$ in the BHM design due to MCMC sampling, and above the threshold in all other patterns. High error rates are observed with few small inactive baskets (Ascending pattern), which decrease as the number of inactive baskets increases (BGN pattern). As before, Fujikawa's design and the power prior design with CPP weights are particularly striking. Looking at the Ascending pattern, Fujikawa's design records a rate of $61 \%$ and CPP of $49 \%$, while the rates of the other models are between $24 \%$ and $30 \%$. In the BGN pattern, Fujikawa's design and the power prior design with CPP weights have error rates of $35$ and $30$ percent, while the values of the other designs are at or below $20 \%$. When the inactive baskets have larger sample sizes (Descending pattern), the TOERs decrease further, and a higher number of inactive baskets (SGN pattern) leads to an additional improvement. 

\noindent The results regarding bias are listed in the \nameref{Supplementary Material} .

\section{Discussion}
\label{sec:discussion} 
\noindent Basket trials are an important part of oncology research, where the sample sizes of individual baskets usually differ. Neglecting the varying sample sizes risks overlapping information from small baskets with information from larger baskets in the borrowing process. This problem can be avoided by limiting the amount of information shared based on the size of the baskets. One way to do this is to use the APP or the new LCPP weights in the power prior design. 

\noindent We compared the power prior design with APP and LCPP weights, containing a quantity limiting parameter, to the CPP weights, which do not limit the amount of information shared from larger baskets, and to several designs from the literature, namely the design proposed by \textcite{fujikawa2020} and the designs BMA, BHM, and EXNEX. All methods were individually optimized and subsequently compared in different scenarios with five baskets of varying sizes. 

\noindent To accurately access the effect of a treatment in clinical practice, a precise overall evaluation of the design used is essential. This includes correctly classifying the baskets as active if the treatment actually has an effect and as inactive if the expected effect is absent, which is represented by the parameter ECD. However, since in a real clinical trial it is not known which and how many of the baskets will respond to treatment, our main focus was the mean number of correctly classified baskets, averaged over all patterns of potential response rates.

\noindent It should be noted that all designs have been tuned according to their maximum mean ECD averaged over all patterns, thus results may vary within individual patterns and baskets.

\noindent According to \textcite{psioda2021}, the BMA design attempts to strike a balance in its performance between the different patterns, which was confirmed in our simulation study. Compared to the other methods, the design achieved rather mediocre results in terms of TOER and power. Looking at the mean ECD, the design performed poorly, always recording the fewest or second fewest correct decisions.

\noindent Due to their similarity, which was enhanced by the tuning process, the BHM and EXNEX designs showed very similar results overall. The similar performance has already been reported by \textcite{daniells2023} and \textcite{baumann2024}. Considering the power, both designs had a rather poor performance, which was especially bad for the SGN pattern. However, compared to the other models, they recorded very low TOERs in the small baskets. The mean ECDs observed were also rather mediocre. \textcite{berry2013} reported good results in terms of the number of correct decisions, which was not confirmed by our study. However, it is important to note that they considered different models in their comparison and that the majority of the designs considered in this paper were not known at that time.

\noindent The simulation study revealed a high sensitivity to varying sample sizes in Fujikawa's design and the power prior design using CPP weights, which was not observed in the simulation study conducted by \textcite{baumann2024} due to equal-sized baskets. Both designs recorded extremely high TOERs in small inactive baskets, which were particularly high when only a few baskets showed no activity (Ascending pattern). The error rates of Fujikawa's design exceeded those of the power prior design with CPP weights. Looking at the FWER, the same trend emerged: high error rates for small and especially for few inactive baskets. The power was less affected by the changing sample sizes. Both designs achieved very good values in the small active baskets, which were noticeably higher than those of the other designs. However, they showed rather low power values in the large baskets. In terms of mean ECD, Fujikawa's design showed poor results, while the power prior design with CPP weights performed quite well.

\noindent The use of APP weights, which have not been applied in a basket trial setting before, or LCPP weights, which we developed and proposed in this work, resulted in robustness to varying sample sizes. In the model comparison, both designs achieved very low TOERs in the small baskets of the Ascending and BGN patterns. However, the designs did not always have low error rates. In the Descending pattern, which has few large inactive baskets, both designs had quite high TOERs compared to the other models. Which method performed better varied between patterns and baskets. This was also present in terms of power. While the APP weights performed rather average in the small baskets of the heterogeneous patterns, the LCPP weights achieved good power values. In most cases, the latter reached the highest power in the model comparison, with the BGN pattern being the only exception with a poor performance. Both designs showed good results in terms of mean ECD, exceeding those of the CPP weights.

\noindent In summary, the three versions of the power prior design achieved the highest mean ECD in all three sample size scenarios. Restricting the amount of shared information, as is the case with the APP and LCPP weights, improved the reliability with which the models classified baskets as active or\break inactive. 

\noindent Overall, the power prior design using LCPP weights was found to be the most appropriate for basket trials with unequal sample sizes in the simulation study conducted. By combining the quantity limiting parameter with the CPP weights, which enable tuning with two parameters, it is robust to varying sample sizes and also best detects both an effect of the investigated treatment and its absence in the individual baskets.

\noindent However, it should be noted that the APP weights performed almost as well in the design comparison, even though they do not include tuning parameters and are therefore easier in application than the other models. 

\noindent Even though, using LCPP weights resulted in the best performance in the present simulation study, there were still fluctuations in the operating characteristics. To control these and further optimize the approach, future research could investigate different weight calculations and also alternative ways to limit the total amount of information to be shared. 
Furthermore, to focus on the effect of borrowing, we only considered a single stage design and did not take into account potential interim analyses that would affect the sample size. These could also be included in future work.

\newpage

\printbibliography[title=References]

\part*{Supplementary Material}
\label{Supplementary Material}
\addcontentsline{toc}{part}{Appendix}
\appendix

\section{Rejection Rates}

\begin{table}[H]
\centering
   \caption[Rejection Rates and FWERs for the linear Scenario.]{Basket-wise rejection rates and FWERs for the linear scenario. For each pattern, results are shown for all designs. Within patterns, the active baskets are highlighted in bold.}
{\small
\begin{tabular}{lccccccc}
  \toprule
\multirow{2}{*}{Pattern} & \multirow{2}{*}{Design} & Basket 1 & Basket 2 & Basket 3 & Basket 4 & Basket 5 & \multirow{2}{*}{FWER} \\ 
 & & $n_1 = 10$ & $n_2 = 15$ & $n_3 = 20$ & $n_4 = 25$ & $n_5 = 30$ & \\
  \hline
\multirow{7}{*}{Null}  
                      & CPP & 0.022 & 0.017 & 0.016 & 0.016 & 0.015 & 0.050 \\ 
                      & APP & 0.008 & 0.014 & 0.017 & 0.018 & 0.018 & 0.048 \\ 
                      & LCPP & 0.014 & 0.015 & 0.015 & 0.017 & 0.018 & 0.050 \\ 
                      & Fujikawa & 0.018 & 0.020 & 0.018 & 0.019 & 0.016 & 0.048 \\
                      & BMA & 0.011 & 0.015 & 0.018 & 0.020 & 0.021 & 0.049 \\ 
                      & BHM & 0.008 & 0.013 & 0.014 & 0.015 & 0.017 & 0.048 \\ 
                      & EXNEX & 0.008 & 0.013 & 0.015 & 0.017 & 0.019 & 0.051 \\ 
   \hline
 \multirow{7}{*}{Alternative}  
                              & CPP & \textbf{0.888} & \textbf{0.922} & \textbf{0.937} & \textbf{0.946} & \textbf{0.951} & 0.000 \\ 
                              & APP & \textbf{0.844} & \textbf{0.910} & \textbf{0.939} & \textbf{0.956} & \textbf{0.968} & 0.000 \\ 
                              & LCPP & \textbf{0.834} & \textbf{0.914} & \textbf{0.951} & \textbf{0.963} & \textbf{0.970} & 0.000 \\ 
                              & Fujikawa & \textbf{0.900} & \textbf{0.923} &\textbf{ 0.929} & \textbf{0.933} & \textbf{0.930} & 0.000 \\
                              & BMA & \textbf{0.851} & \textbf{0.896} & \textbf{0.917} & \textbf{0.933} & \textbf{0.949} & 0.000 \\ 
                              & BHM & \textbf{0.845} & \textbf{0.890} & \textbf{0.920} & \textbf{0.936} & \textbf{0.955} & 0.000 \\ 
                              & EXNEX & \textbf{0.843} & \textbf{0.891} & \textbf{0.920} & \textbf{0.937} & \textbf{0.955} & 0.000 \\ 
    \hline
 \multirow{7}{*}{Ascending}  
                            & CPP & 0.313 & 0.297 & \textbf{0.607} & \textbf{0.860} & \textbf{0.869} & 0.461 \\ 
                            & APP & 0.164 & 0.214 & \textbf{0.553} & \textbf{0.850} & \textbf{0.880} & 0.297 \\ 
                            & LCPP & 0.164 & 0.233 & \textbf{0.590} & \textbf{0.866} & \textbf{0.890} & 0.314 \\ 
                            & Fujikawa & 0.371 & 0.310 & \textbf{0.598} & \textbf{0.853} & \textbf{0.857} & 0.523 \\
                            & BMA & 0.191 & 0.196 & \textbf{0.483} & \textbf{0.830} & \textbf{0.863} & 0.271 \\ 
                            & BHM & 0.172 & 0.180 & \textbf{0.465} & \textbf{0.813} & \textbf{0.856} & 0.264 \\ 
                            & EXNEX & 0.178 & 0.182 & \textbf{0.489} & \textbf{0.824} & \textbf{0.867} & 0.273 \\ 
    \hline
 \multirow{7}{*}{Descending} 
                             & CPP & \textbf{0.539} & \textbf{0.553} & \textbf{0.317} & 0.095 & 0.081 & 0.133 \\ 
                             & APP & \textbf{0.456} & \textbf{0.578} & \textbf{0.383} & 0.132 & 0.123 & 0.198 \\ 
                             & LCPP & \textbf{0.497} & \textbf{0.584} & \textbf{0.367} & 0.132 & 0.117 & 0.181 \\ 
                             & Fujikawa & \textbf{0.544} & \textbf{0.582} & \textbf{0.361} & 0.105 & 0.081 & 0.152 \\ 
                             & BMA & \textbf{0.401} &\textbf{ 0.500} & \textbf{0.306} & 0.106 & 0.104 & 0.154 \\ 
                             & BHM & \textbf{0.395} & \textbf{0.512} & \textbf{0.327} & 0.096 & 0.098 & 0.163 \\ 
                             & EXNEX & \textbf{0.397} & \textbf{0.513} & \textbf{0.326} & 0.098 & 0.098 & 0.162 \\ 
    \hline
 \multirow{7}{*}{BGN}  
                      & CPP & 0.138 & 0.110 & 0.094 & 0.089 & \textbf{0.788} & 0.280 \\ 
                      & APP & 0.056 & 0.075 & 0.080 & 0.081 & \textbf{0.801} & 0.197 \\ 
                      & LCPP & 0.059 & 0.074 & 0.084 & 0.097 & \textbf{0.789} & 0.207 \\ 
                      & Fujikawa & 0.163 & 0.126 & 0.099 & 0.082 & \textbf{0.787} & 0.299 \\
                      & BMA & 0.054 & 0.058 & 0.064 & 0.066 & \textbf{0.785} & 0.165 \\ 
                      & BHM & 0.048 & 0.054 & 0.058 & 0.063 & \textbf{0.812} & 0.160 \\ 
                      & EXNEX & 0.049 & 0.058 & 0.061 & 0.067 &\textbf{0.818} & 0.169 \\   
    \hline
 \multirow{7}{*}{SGN}  
                      & CPP & \textbf{0.418} & 0.037 & 0.029 & 0.027 & 0.026 & 0.074 \\ 
                      & APP & \textbf{0.271} & 0.039 & 0.039 & 0.044 & 0.040 & 0.108 \\ 
                      & LCPP & \textbf{0.397} & 0.040 & 0.033 & 0.034 & 0.032 & 0.088 \\ 
                      & Fujikawa & \textbf{0.319} & 0.045 & 0.037 & 0.033 & 0.029 & 0.090 \\
                      & BMA & \textbf{0.260} & 0.035 & 0.035 & 0.039 & 0.040 & 0.093 \\ 
                      & BHM & \textbf{0.246} & 0.030 & 0.034 & 0.039 & 0.040 & 0.106 \\ 
                      & EXNEX & \textbf{0.250} & 0.029 & 0.034 & 0.039 & 0.042 & 0.107 \\
   \bottomrule
   \end{tabular}
   }
   \label{tab:rr_linear}
\end{table}

\begin{table}[H]
\centering
   \caption[Rejection Rates and FWERs for the High Variance Scenario.]{Basket-wise rejection rates and FWERs for the scenario with a high variance. For each pattern, results are shown for all designs. Within patterns, the active baskets are highlighted in bold.}
{\small
\begin{tabular}{lccccccc}
  \toprule
\multirow{2}{*}{Pattern} & \multirow{2}{*}{Design} & Basket 1 & Basket 2 & Basket 3 & Basket 4 & Basket 5 & \multirow{2}{*}{FWER} \\ 
& & $n_1 = 10$ & $n_2 = 10$ & $n_3 = 10$ & $n_4 = 20$ & $n_5 = 50$ & \\
  \hline
\multirow{7}{*}{Null}  
                      & CPP & 0.018 & 0.018 & 0.017 & 0.013 & 0.014 & 0.048 \\ 
                      & APP & 0.013 & 0.014 & 0.012 & 0.014 & 0.018 & 0.048 \\ 
                      & LCPP & 0.017 & 0.015 & 0.015 & 0.015 & 0.020 & 0.050 \\ 
                      & Fujikawa & 0.020 & 0.021 & 0.019 & 0.019 & 0.010 & 0.050 \\
                      & BMA & 0.015 & 0.015 & 0.013 & 0.018 & 0.028 & 0.050 \\ 
                      & BHM & 0.010 & 0.010 & 0.009 & 0.015 & 0.026 & 0.050 \\ 
                      & EXNEX & 0.009 & 0.009 & 0.009 & 0.015 & 0.028 & 0.050 \\  
   \hline
 \multirow{7}{*}{Alternative}  
                              & CPP & \textbf{0.866} & \textbf{0.865} & \textbf{0.868} & \textbf{0.905} & \textbf{0.954} & 0.000 \\ 
                              & APP & \textbf{0.825} & \textbf{0.827} & \textbf{0.827} & \textbf{0.919} & \textbf{0.988} & 0.000 \\ 
                              & LCPP & \textbf{0.871} & \textbf{0.870} & \textbf{0.872} & \textbf{0.954} & \textbf{0.989} & 0.000 \\ 
                              & Fujikawa & \textbf{0.863} & \textbf{0.860} & \textbf{0.862} & \textbf{0.864} & \textbf{0.918} & 0.000 \\
                              & BMA & \textbf{0.845} & \textbf{0.845} & \textbf{0.847} & \textbf{0.908} & \textbf{0.986} & 0.000 \\ 
                              & BHM & \textbf{0.838} & \textbf{0.841} & \textbf{0.842} & \textbf{0.911} & \textbf{0.989} & 0.000 \\ 
                              & EXNEX & \textbf{0.830} & \textbf{0.829} & \textbf{0.831} & \textbf{0.905} & \textbf{0.988} & 0.000 \\ 
    \hline
 \multirow{7}{*}{Ascending} 
                            & CPP & 0.287 & 0.289 & \textbf{0.533} & \textbf{0.845} & \textbf{0.922} & 0.434 \\ 
                            & APP & 0.168 & 0.166 & \textbf{0.354} & \textbf{0.769} & \textbf{0.949} & 0.251 \\ 
                            & LCPP & 0.216 & 0.215 & \textbf{0.370} & \textbf{0.802} & \textbf{0.957} & 0.302 \\ 
                            & Fujikawa & 0.307 & 0.307 & \textbf{0.547} & \textbf{0.823} & \textbf{0.891} & 0.465 \\ 
                            & BMA & 0.201 & 0.204 & \textbf{0.403} & \textbf{0.799} & \textbf{0.962} & 0.287 \\ 
                            & BHM & 0.191 & 0.190 & \textbf{0.381} & \textbf{0.764} & \textbf{0.962} & 0.279 \\ 
                            & EXNEX & 0.196 & 0.198 & \textbf{0.406} & \textbf{0.785} & \textbf{0.963} & 0.300 \\ 
    \hline
 \multirow{7}{*}{Descending}  
                             & CPP & \textbf{0.406} & \textbf{0.415} & \textbf{0.228} & 0.048 & 0.027 & 0.061 \\ 
                             & APP & \textbf{0.452} & \textbf{0.456} & \textbf{0.298} & 0.095 & 0.080 & 0.146 \\ 
                             & LCPP & \textbf{0.491} & \textbf{0.495} & \textbf{0.357} & 0.122 & 0.078 & 0.157 \\ 
                             & Fujikawa & \textbf{0.460} & \textbf{0.464} & \textbf{0.292} & 0.060 & 0.017 & 0.070 \\
                             & BMA & \textbf{0.321} & \textbf{0.330} & \textbf{0.176} & 0.073 & 0.085 & 0.117 \\ 
                             & BHM & \textbf{0.325} & \textbf{0.339} & \textbf{0.186} & 0.074 & 0.091 & 0.139 \\ 
                             & EXNEX & \textbf{0.320} & \textbf{0.330} & \textbf{0.177} & 0.069 & 0.089 & 0.134 \\ 
    \hline
 \multirow{7}{*}{BGN} 
                      & CPP & 0.173 & 0.173 & 0.172 & 0.131 & \textbf{0.959} & 0.456 \\ 
                      & APP & 0.070 & 0.070 & 0.068 & 0.090 & \textbf{0.959} & 0.189 \\ 
                      & LCPP & 0.070 & 0.072 & 0.070 & 0.123 & \textbf{0.954} & 0.197 \\ 
                      & Fujikawa & 0.186 & 0.184 & 0.180 & 0.112 & \textbf{0.952} & 0.460 \\ 
                      & BMA & 0.075 & 0.078 & 0.075 & 0.081 & \textbf{0.969} & 0.217 \\ 
                      & BHM & 0.069 & 0.067 & 0.064 & 0.077 & \textbf{0.974} & 0.184 \\ 
                      & EXNEX & 0.072 & 0.072 & 0.070 & 0.086 & \textbf{0.974} & 0.200 \\   
    \hline
 \multirow{7}{*}{SGN}  
                      & CPP & \textbf{0.377} & 0.043 & 0.043 & 0.029 & 0.017 & 0.096 \\ 
                      & APP & \textbf{0.303} & 0.050 & 0.053 & 0.039 & 0.032 & 0.127 \\ 
                      & LCPP & \textbf{0.339} & 0.065 & 0.066 & 0.044 & 0.033 & 0.134 \\ 
                      & Fujikawa & \textbf{0.343} & 0.057 & 0.058 & 0.036 & 0.013 & 0.117 \\
                      & BMA & \textbf{0.267} & 0.029 & 0.031 & 0.041 & 0.045 & 0.093 \\ 
                      & BHM & \textbf{0.258} & 0.024 & 0.028 & 0.039 & 0.048 & 0.106 \\ 
                      & EXNEX & \textbf{0.265} & 0.023 & 0.026 & 0.038 & 0.047 & 0.103 \\ 
   \bottomrule
   \end{tabular}
   }
   \label{tab:rr_highVar}
\end{table}

\newpage
\section{Bias}
\begin{table}[H]
\centering
   \caption[Bias for the Linear Scenario.]{Bias for the linear scenario. For each pattern, results are shown for all designs.}
{\small
\begin{tabular}{lcccccc}
  \toprule
\multirow{2}{*}{Pattern} & \multirow{2}{*}{Design} & Basket 1 & Basket 2 & Basket 3 & Basket 4 & Basket 5  \\ 
 & & $n_1 = 10$ & $n_2 = 15$ & $n_3 = 20$ & $n_4 = 25$ & $n_5 = 30$  \\
  \hline
\multirow{7}{*}{Null}  
                      & CPP & 0.011 & 0.008 & 0.007 & 0.007 & 0.006 \\ 
                      & APP & 0.018 & 0.013 & 0.011 & 0.010 & 0.009 \\ 
                      & LCPP & 0.017 & 0.011 & 0.008 & 0.007 & 0.006 \\ 
                      & Fujikawa & 0.040 & 0.034 & 0.031 & 0.030 & 0.028 \\
                      & BMA & 0.007 & 0.006 & 0.005 & 0.005 & 0.005 \\ 
                      & BHM & 0.002 & 0.001 & 0.000 & 0.000 & 0.001 \\ 
                      & EXNEX & 0.001 & 0.001 & 0.000 & 0.001 & 0.000 \\ 
   \hline
 \multirow{7}{*}{Alternative}  
                              & CPP & 0.005 & 0.006 & 0.004 & 0.003 & 0.003 \\ 
                              & APP & 0.007 & 0.008 & 0.006 & 0.004 & 0.004 \\ 
                              & LCPP & 0.008 & 0.007 & 0.005 & 0.003 & 0.003 \\ 
                              & Fujikawa & 0.016 & 0.017 & 0.014 & 0.013 & 0.012 \\
                              & BMA & 0.003 & 0.004 & 0.003 & 0.002 & 0.002 \\ 
                              & BHM & 0.001 & 0.002 & 0.001 & 0.000 & 0.000 \\ 
                              & EXNEX & 0.001 & 0.002 & 0.001 & 0.000 & 0.000 \\ 
    \hline
 \multirow{7}{*}{Ascending} 
                            & CPP & 0.076 & 0.075 & 0.023 & 0.024 & 0.023 \\ 
                            & APP & 0.081 & 0.074 & 0.018 & 0.033 & 0.032 \\ 
                            & LCPP & 0.070 & 0.072 & 0.018 & 0.032 & 0.030 \\ 
                            & Fujikawa & 0.100 & 0.087 & 0.036 & 0.011 & 0.011 \\ 
                            & BMA & 0.082 & 0.074 & 0.020 & 0.034 & 0.033 \\ 
                            & BHM & 0.082 & 0.073 & 0.010 & 0.038 & 0.037 \\ 
                            & EXNEX & 0.071 & 0.069 & 0.014 & 0.036 & 0.035 \\
    \hline
 \multirow{7}{*}{Descending}  
                             & CPP & 0.052 & 0.055 & 0.017 & 0.037 & 0.034 \\ 
                             & APP & 0.053 & 0.052 & 0.001 & 0.048 & 0.045 \\ 
                             & LCPP & 0.039 & 0.050 & 0.011 & 0.045 & 0.040 \\ 
                             & Fujikawa & 0.031 & 0.028 & 0.011 & 0.049 & 0.045 \\
                             & BMA & 0.065 & 0.060 & 0.008 & 0.041 & 0.039 \\ 
                             & BHM & 0.075 & 0.068 & 0.012 & 0.039 & 0.036 \\ 
                             & EXNEX & 0.075 & 0.067 & 0.012 & 0.038 & 0.036 \\
    \hline
 \multirow{7}{*}{BGN}  
                      & CPP & 0.034 & 0.030 & 0.029 & 0.027 & 0.051 \\ 
                      & APP & 0.044 & 0.036 & 0.032 & 0.029 & 0.071 \\ 
                      & LCPP & 0.035 & 0.030 & 0.029 & 0.031 & 0.071 \\ 
                      & Fujikawa & 0.065 & 0.054 & 0.046 & 0.041 & 0.030 \\
                      & BMA & 0.045 & 0.037 & 0.032 & 0.028 & 0.062 \\ 
                      & BHM & 0.038 & 0.032 & 0.028 & 0.025 & 0.069 \\ 
                      & EXNEX & 0.033 & 0.030 & 0.029 & 0.027 & 0.068 \\  
    \hline
 \multirow{7}{*}{SGN} 
                      & CPP & 0.089 & 0.016 & 0.013 & 0.013 & 0.011 \\ 
                      & APP & 0.112 & 0.030 & 0.024 & 0.022 & 0.020 \\ 
                      & LCPP & 0.081 & 0.024 & 0.017 & 0.015 & 0.013 \\ 
                      & Fujikawa & 0.060 & 0.041 & 0.035 & 0.033 & 0.030 \\ 
                      & BMA & 0.118 & 0.023 & 0.019 & 0.018 & 0.016 \\ 
                      & BHM & 0.141 & 0.020 & 0.016 & 0.015 & 0.013 \\ 
                      & EXNEX & 0.139 & 0.017 & 0.016 & 0.016 & 0.014 \\
   \bottomrule
   \end{tabular}
   }
   \label{tab:bias_linear}
\end{table}

\begin{table}[H]
\centering
   \caption[Bias for the Grouped Scenario.]{Bias for the grouped scenario. For each pattern, results are shown for all designs.}
{\small
\begin{tabular}{lcccccc}
  \toprule
\multirow{2}{*}{Pattern} & \multirow{2}{*}{Design} & Basket 1 & Basket 2 & Basket 3 & Basket 4 & Basket 5  \\ 
 & & $n_1 = 10$ & $n_2 = 10$ & $n_3 = 25$ & $n_4 = 25$ & $n_5 = 30$  \\
  \hline
\multirow{7}{*}{Null}  
                      & CPP & 0.011 & 0.010 & 0.006 & 0.006 & 0.005 \\ 
                      & APP & 0.018 & 0.017 & 0.009 & 0.010 & 0.008 \\ 
                      & LCPP & 0.017 & 0.016 & 0.006 & 0.006 & 0.005 \\ 
                      & Fujikawa & 0.038 & 0.037 & 0.029 & 0.030 & 0.027 \\
                      & BMA & 0.007 & 0.006 & 0.004 & 0.005 & 0.004 \\ 
                      & BHM & 0.002 & 0.001 & 0.001 & 0.000 & 0.001 \\ 
                      & EXNEX & 0.001 & 0.002 & 0.000 & 0.001 & 0.000 \\ 
   \hline
 \multirow{7}{*}{Alternative} 
                              & CPP & 0.006 & 0.007 & 0.003 & 0.003 & 0.002 \\ 
                              & APP & 0.007 & 0.008 & 0.004 & 0.004 & 0.003 \\ 
                              & LCPP & 0.008 & 0.009 & 0.003 & 0.002 & 0.002 \\ 
                              & Fujikawa & 0.016 & 0.017 & 0.012 & 0.012 & 0.011 \\ 
                              & BMA & 0.003 & 0.003 & 0.002 & 0.002 & 0.001 \\ 
                              & BHM & 0.001 & 0.001 & 0.000 & 0.001 & 0.001 \\ 
                              & EXNEX & 0.000 & 0.000 & 0.000 & 0.000 & 0.001 \\ 
    \hline
 \multirow{7}{*}{Ascending}  
                            & CPP & 0.079 & 0.079 & 0.026 & 0.024 & 0.023 \\ 
                            & APP & 0.081 & 0.081 & 0.020 & 0.033 & 0.032 \\ 
                            & LCPP & 0.072 & 0.071 & 0.027 & 0.035 & 0.032 \\
                            & Fujikawa & 0.112 & 0.112 & 0.036 & 0.016 & 0.015 \\
                            & BMA & 0.085 & 0.084 & 0.021 & 0.035 & 0.034 \\ 
                            & BHM & 0.085 & 0.085 & 0.010 & 0.038 & 0.036 \\ 
                            & EXNEX & 0.074 & 0.073 & 0.014 & 0.036 & 0.034 \\
    \hline
 \multirow{7}{*}{Descending}  
                             & CPP & 0.056 & 0.055 & 0.022 & 0.037 & 0.035 \\ 
                             & APP & 0.054 & 0.053 & 0.006 & 0.048 & 0.044 \\ 
                             & LCPP & 0.042 & 0.042 & 0.025 & 0.046 & 0.042 \\
                             & Fujikawa & 0.048 & 0.047 & 0.005 & 0.053 & 0.049 \\
                             & BMA & 0.069 & 0.068 & 0.012 & 0.041 & 0.038 \\ 
                             & BHM & 0.079 & 0.079 & 0.014 & 0.038 & 0.035 \\ 
                             & EXNEX & 0.079 & 0.078 & 0.014 & 0.037 & 0.035 \\ 
    \hline
 \multirow{7}{*}{BGN}  
                      & CPP & 0.035 & 0.034 & 0.028 & 0.027 & 0.052 \\ 
                      & APP & 0.044 & 0.044 & 0.030 & 0.029 & 0.071 \\ 
                      & LCPP & 0.036 & 0.036 & 0.033 & 0.033 & 0.078 \\ 
                      & Fujikawa & 0.070 & 0.069 & 0.047 & 0.046 & 0.044 \\
                      & BMA & 0.046 & 0.045 & 0.029 & 0.028 & 0.062 \\ 
                      & BHM & 0.038 & 0.038 & 0.026 & 0.026 & 0.069 \\ 
                      & EXNEX & 0.033 & 0.033 & 0.028 & 0.027 & 0.068 \\    
    \hline
 \multirow{7}{*}{SGN} 
                      & CPP & 0.089 & 0.021 & 0.012 & 0.012 & 0.012 \\ 
                      & APP & 0.111 & 0.040 & 0.022 & 0.021 & 0.021 \\ 
                      & LCPP & 0.088 & 0.040 & 0.015 & 0.015 & 0.014 \\ 
                      & Fujikawa & 0.094 & 0.049 & 0.035 & 0.034 & 0.033 \\ 
                      & BMA & 0.118 & 0.028 & 0.018 & 0.017 & 0.017 \\ 
                      & BHM & 0.141 & 0.023 & 0.015 & 0.015 & 0.014 \\ 
                      & EXNEX & 0.138 & 0.018 & 0.015 & 0.015 & 0.015 \\
   \bottomrule
   \end{tabular}
   }
   \label{tab:bias_grouped}
\end{table}

\begin{table}[H]
\centering
   \caption[Bias for the High Variance Scenario.]{Bias for the scenario with a high variance. For each pattern, results are shown for all designs.}
{\small
\begin{tabular}{lcccccc}
  \toprule
\multirow{2}{*}{Pattern} & \multirow{2}{*}{Design} & Basket 1 & Basket 2 & Basket 3 & Basket 4 & Basket 5  \\ 
 & & $n_1 = 10$ & $n_2 = 10$ & $n_3 = 10$ & $n_4 = 20$ & $n_5 = 50$  \\
  \hline
\multirow{7}{*}{Null}  
                      & CPP & 0.011 & 0.011 & 0.012 & 0.008 & 0.005 \\ 
                      & APP & 0.017 & 0.017 & 0.017 & 0.012 & 0.008 \\ 
                      & LCPP & 0.014 & 0.014 & 0.015 & 0.009 & 0.005 \\ 
                      & Fujikawa & 0.045 & 0.045 & 0.046 & 0.032 & 0.020 \\
                      & BMA & 0.007 & 0.007 & 0.007 & 0.005 & 0.004 \\ 
                      & BHM & 0.002 & 0.002 & 0.002 & 0.000 & 0.001 \\ 
                      & EXNEX & 0.003 & 0.003 & 0.003 & 0.001 & 0.001 \\  
   \hline
 \multirow{7}{*}{Alternative}  
                              & CPP & 0.006 & 0.006 & 0.007 & 0.003 & 0.004 \\ 
                              & APP & 0.007 & 0.007 & 0.008 & 0.004 & 0.004 \\ 
                              & LCPP & 0.007 & 0.007 & 0.007 & 0.003 & 0.003 \\ 
                              & Fujikawa & 0.019 & 0.019 & 0.019 & 0.012 & 0.010 \\
                              & BMA & 0.003 & 0.003 & 0.003 & 0.002 & 0.002 \\ 
                              & BHM & 0.001 & 0.001 & 0.001 & 0.001 & 0.000 \\ 
                              & EXNEX & 0.000 & 0.001 & 0.000 & 0.000 & 0.001 \\ 
    \hline
 \multirow{7}{*}{Ascending}  
                            & CPP & 0.078 & 0.077 & 0.028 & 0.015 & 0.007 \\ 
                            & APP & 0.080 & 0.080 & 0.016 & 0.034 & 0.022 \\ 
                            & LCPP & 0.078 & 0.078 & 0.005 & 0.039 & 0.020 \\ 
                            & Fujikawa & 0.088 & 0.087 & 0.046 & 0.001 & 0.001 \\
                            & BMA & 0.086 & 0.086 & 0.027 & 0.030 & 0.021 \\ 
                            & BHM & 0.088 & 0.088 & 0.019 & 0.037 & 0.025 \\ 
                            & EXNEX & 0.073 & 0.072 & 0.023 & 0.032 & 0.022 \\ 
    \hline
 \multirow{7}{*}{Descending}  
                             & CPP & 0.054 & 0.052 & 0.014 & 0.026 & 0.017 \\ 
                             & APP & 0.054 & 0.052 & 0.008 & 0.050 & 0.032 \\ 
                             & LCPP & 0.054 & 0.052 & 0.011 & 0.050 & 0.026 \\ 
                             & Fujikawa & 0.011 & 0.009 & 0.025 & 0.042 & 0.022 \\
                             & BMA & 0.068 & 0.066 & 0.012 & 0.034 & 0.024 \\ 
                             & BHM & 0.083 & 0.082 & 0.019 & 0.035 & 0.022 \\ 
                             &EXNEX & 0.082 & 0.080 & 0.021 & 0.033 & 0.023 \\  
    \hline
 \multirow{7}{*}{BGN}  
                      & CPP & 0.044 & 0.044 & 0.043 & 0.037 & 0.017 \\ 
                      & APP & 0.045 & 0.045 & 0.044 & 0.039 & 0.050 \\ 
                      & LCPP & 0.035 & 0.035 & 0.035 & 0.039 & 0.047 \\ 
                      & Fujikawa & 0.068 & 0.068 & 0.067 & 0.046 & 0.005 \\
                      & BMA & 0.054 & 0.054 & 0.053 & 0.038 & 0.038 \\ 
                      & BHM & 0.046 & 0.046 & 0.046 & 0.035 & 0.042 \\ 
                      & EXNEX & 0.041 & 0.041 & 0.040 & 0.040 & 0.041 \\    
    \hline
 \multirow{7}{*}{SGN} 
                      & CPP & 0.079 & 0.021 & 0.021 & 0.014 & 0.009 \\ 
                      & APP & 0.108 & 0.040 & 0.040 & 0.028 & 0.018 \\ 
                      & LCPP & 0.111 & 0.042 & 0.042 & 0.025 & 0.012 \\ 
                      & Fujikawa & 0.027 & 0.052 & 0.052 & 0.035 & 0.019 \\ 
                      & BMA & 0.115 & 0.028 & 0.028 & 0.019 & 0.013 \\ 
                      & BHM & 0.138 & 0.024 & 0.024 & 0.018 & 0.010 \\ 
                      & EXNEX & 0.132 & 0.016 & 0.016 & 0.018 & 0.012 \\
   \bottomrule
   \end{tabular}
   }
   \label{tab:bias_HighVar}
\end{table}

\end{document}